# Kagome−Triangular Lattice Antiferromagnet NaBa$_2$Mn$_3$F$_{11}$

Hajime Ishikawa*, Tsuyoshi Okubo, Yoshihiko Okamoto, and Zenji Hiroi

*Institute for Solid State Physics, University of Tokyo, Kashiwanoha 5-1-5, Kashiwa, Chiba 277-8581, Japan*

A new type of magnetically frustrated lattice is found in the layered fluoride NaBa$_2$Mn$_3$F$_{11}$. A kagome-type array of regular triangles composed of Mn$^{2+}$ ions (spin 5/2) deforms so as to generate the next-nearest-neighbor interaction $J_2$ between three out of six spins in the hexagon of a normal kagome lattice, in addition to the nearest-neighbor interaction $J_1$. As a function of $|J_2/J_1|$, this lattice can interconnect the kagome ($J_2 = 0$) and the triangular ($J_2 = J_1$) lattices and thus is called the kagome−triangular (KT) lattice. Magnetic susceptibility and heat capacity measurements performed on a polycrystalline sample of NaBa$_2$Mn$_3$F$_{11}$ show an intensive short-range antiferromagnetic correlation below 14 K probably due to the specific magnetic frustration of the KT lattice. In addition, a long-range order at 2.0 K is observed, which is significantly low compared with the antiferromagnetic Weiss temperature of 32.3 K. Theoretical considerations of Heisenberg spins in the KT lattice reveal unique non-coplanar magnetic orders in the case of ferromagnetic $J_1$ and antiferromagnetic $J_2$. NaBa$_2$Mn$_3$F$_{11}$ may actually exist in this regime according to the results of our analysis based on classical Monte−Carlo simulation.

*E-mail: ishikawa@issp.u-tokyo.ac.jp

Frustrated magnets have been extensively studied in the field of magnetism. Typical examples are Heisenberg antiferromagnets with two-dimensional triangular and kagome lattices. In the former, theoretical studies reveal that ordered ground states with 120º spin arrangements are selected to reach a compromise with the geometrical frustration,[1,2] which have been found experimentally in real materials such as Ba$_3$CoSb$_2$O$_9$ ($S$ = 1/2),[3] LiCrO$_2$ ($S$ = 3/2),[4] and Rb$_4$Mn(MoO$_4$)$_3$ ($S$ = 5/2).[5] In contrast, the classical ground states of Heisenberg kagome antiferromagnets can macroscopically degenerate: it is believed that a set of coplanar states is selected by the order-by-disorder mechanism.[6-8] For quantum spins on the kagome lattice, exotic ground states such as spin liquids or valence-bond crystal states may be stable.[9-15] Although no magnetic long-range orders are expected for the pure Heisenberg kagome model with a nearest-neighbor coupling $J_1$, some perturbations, such as a next-nearest-neighbor interaction, a Dzyaloshinskii–Moriya (DM) interaction, or the deformation of the lattice, may drive the system toward a certain long-range order. For example, the infinitesimal next-nearest-neighbor coupling $J_2$ stabilizes coplanar orders with 120º spin arrangements:[16] the $q = 0$ order is favored for antiferromagnetic $J_2$, while the $\sqrt{3} \times \sqrt{3}$ order for ferromagnetic $J_2$. On the other hand, in the case of ferromagnetic $J_1$ and antiferromagnetic $J_2$, a non-coplanar order with a cuboctahedral symmetry that contains twelve spins per unit cell is predicted.[17-18]

These theoretical predictions for kagome antiferromagnets have been examined experimentally in various compounds. Jarosite [$AM_3(OH)_6(SO_4)_2$; ($A$ = K$^+$, Rb$^+$ etc.; $M$ = Fe$^{3+}$, Cr$^{3+}$)] and SrCr$_8$Ga$_4$O$_{19}$ are typical candidate compounds for classical spin systems, which show transitions to either 120º orders or spin-glass orders.[19-21] For quantum spin systems, many copper minerals have been studied. Herbertsmithite [$ZnCu_3(OH)_6Cl_2$] with a perfect kagome geometry does not show any order down to 50 mK and is considered to realize a spin liquid state.[22-24] Vesignieite [$BaCu_3V_2O_8(OH)_2$] with a small or with no distortion in the kagome lattice show a 120º order with $q = 0$ probably owing to a sizable DM interaction.[25-28] Other copper minerals like volborthite [$Cu_3V_2O_7(OH)_2 \cdot 2H_2O$],[29,30] edwardsite [$Cd_2Cu_3(SO_4)_2(OH)_6 \cdot 4H_2O$],[31] and $KCu_3As_2O_7(OH)_3$[32] all exhibit certain long-range orders depending on the kind of perturbation. In addition, the fluoride Rb$_2$Cu$_3$SnF$_{12}$ with pinwheel-shaped Cu dodecamers shows a valence-bond-solid state.[33]

In this letter, we report the magnetic properties of the layered fluoride NaBa$_2$Mn$_3$F$_{11}$ comprising a new type of frustrated lattice called the kagome–triangular (KT) lattice, which can interconnect the kagome and triangular lattices. NaBa$_2$Mn$_3$F$_{11}$ was synthesized by Darriet et al. in 1992 for the first time.[34] It crystallizes in a trigonal structure of the space group $R\bar{3}c$ with the lattice constants $a$ = 7.003(1) Å and $c$ = 35.466(6) Å. Since the Mn$^{2+}$ ion has the $(3d)^5$ high-spin electronic configuration with a localized spin of $S$ = 5/2, the compound is regarded as a classical Heisenberg spin system. Darriet et al. measured magnetic susceptibility and reported that antiferromagnetic interactions are dominant in NaBa$_2$Mn$_3$F$_{11}$ with a Weiss temperature of 31.5 K.[34] Moreover, no long-range order was observed above 1.8 K. Since then, there had been no reports on the magnetic properties. Thus, we are interested in this compound from the viewpoint of magnetic frustration.

The crystal structure of NaBa$_2$Mn$_3$F$_{11}$ is shown in Fig. 1. All the Mn atoms are crystallographically identical and reside in unique MnF$_7$ pentagonal bipyramids with a [5 + 2] coordination: there are five F atoms in the basal plane and two above and below the Mn atom.[34] The MnF$_7$ pentagonal bipyramids form a Mn$_3$F$_{11}$ layer, where regular triangles composed of Mn atoms are connected to each other by their corners to generate a kagome-type array. The Mn$_3$F$_{11}$ layers are separated from each other by NaF$_8$ and BaF$_9$ coordination polyhedra along the $c$-axis; therefore, there is no direct connection between MnF$_7$ pentagonal bipyramids between adjacent layers. The Mn–Mn distances are 3.562(2) and 5.409(1) Å for the nearest and next-nearest- neighbors in the layer and 6.2586(3) Å between the layers. Thus, one expects a reasonably good two-dimensionality in magnetic interactions. The magnetic interactions between Mn spins in the layer should occur by superexchange interactions via Mn–F–Mn paths. As shown in Fig. 1(c), the nearest-neighbor coupling $J_1$ occurs via the Mn–F1–Mn and Mn–F2–Mn paths, while the next-nearest-neighbor coupling $J_2$ occurs via the Mn–F1–Mn path. Note that, because of the mutual tilting of Mn triangles, $J_2$ coupling exists only on the right side of the pair of Mn triangles in Fig. 1(c).

The KT lattice composed of Mn atoms of NaBa$_2$Mn$_3$F$_{11}$ is schematically shown in Fig. 2(b). The topology of the KT lattice is equivalent to that of the triangular lattice: it becomes a regular triangular lattice for $J_1 = J_2$, as shown in Fig. 2(c). Alternatively, it is generated by eliminating half of the $J_2$ couplings from the conventional $J_1$–$J_2$ kagome lattice [Fig. 2(a)] so that all the remaining $J_2$ triangles inside the hexagons should point to either right or left. Thus, the KT lattice can interconnect the triangular and kagome lattices by changing $|J_2/J_1|$. On one hand, the KT lattice approaches a set of isolated triangular clusters in the limit of $|J_2| \gg |J_1|$. To our knowledge, the KT lattice has not been studied theoretically thus far except for two reports in which the XY model in the KT lattice is studied to interpret the magnetic properties of the ZrNiAl-type compounds mentioned below,[35] and the evolution of the 1/3 magnetization plateau or ramp between the triangular- and kagome-lattice antiferromagnets is examined as a function of $J_2/J_1$.[36]

The KT lattice is rarely found in real compounds, particularly in spin systems. In metallic compounds, on the other hand, it is found in alloys with the ZrNiAl-type structure.[35] For example, Ce$^{3+}$ ions form this type of lattice in CePdAl, where Ce Ising spins show a partially ordered state at low temperature possibly as a result of a competition between a magnetic frustration and a Kondo effect.[37-39] However, there must be a strong three-dimensional character in this alloy, because the in-plane and interplane Ce–Ce distances are similar, i.e.,

3.7 and 4.2 Å, respectively. Thus, it may not be reasonable to assume a two-dimensional KT or kagome lattice of Ce atoms in CePdAl, although the Ce sublattice has been called the distorted kagome lattice or a kagome-like lattice.[35,38-39] In contrast, the Mn sublattice of $NaBa_2Mn_3F_{11}$ apparently takes a good two-dimensional character and thus is called the kagome–triangular lattice in order to distinguish it from other kagome-related lattices with different types of modification.

We prepared a polycrystalline sample of $NaBa_2Mn_3F_{11}$ to study the characteristics of the KT lattice. Magnetic susceptibility and heat capacity measurements reveal that $NaBa_2Mn_3F_{11}$ is an antiferromagnet with a Weiss temperature of 32.3 K and shows a short-range order below 14 K followed by a long-range order at 2.0 K. We also study a classical Heisenberg spin model on the KT lattice and find non-coplanar orders in the $J_1$–$J_2$ phase diagram for ferromagnetic $J_1$ and antiferromagnetic $J_2$. By comparing experimental results with the results of calculation by classical Monte–Carlo simulation, we show that $NaBa_2Mn_3F_{11}$ may exist in this parameter regime for non-coplanar orders. It is demonstrated that $NaBa_2Mn_3F_{11}$ provides us with a unique playground for frustration physics.

A polycrystalline sample of $NaBa_2Mn_3F_{11}$ was synthesized by a solid state reaction. NaF (0.1 g, 99.9%), $BaF_2$ (0.8351 g, 99.9%) and $MnF_2$ (0.6640 g, 99.9%) were ground well in an agate mortar under argon atmosphere and loaded into a Pt tube. The tube was sealed in an evacuated quartz ampoule and heated at 640 °C for 72 h to obtain a white polycrystalline sample. Sample characterization was performed by powder X-ray diffraction analysis, the result of which indicated that a single-phase sample was successfully synthesized. The lattice constants are $a = 7.004(1)$ Å and $c = 35.439(8)$ Å, which are close to those reported previously.[34] Magnetic susceptibility was measured down to 1.8 K in a Magnetic Property Measurement System (Quantum Design), and heat capacity was measured down to 0.4 K in a Physical Property Measurement System (Quantum Design).

The temperature dependences of the magnetic susceptibility $\chi$ and its inverse between 1.8 and 300 K are shown in Fig. 3; the diamagnetic contribution of core electrons, $\chi_0 = -7.73 \times 10^{-5}$ cm$^3$ mol–Mn$^{-1}$, from the literature[40] has already been subtracted. The inverse susceptibility is linear above 100 K following the Curie–Weiss law $\chi = C/(T + \Theta)$, where $C$ and $\Theta$ are the Curie constant and Weiss temperature, respectively. A Curie–Weiss fitting between 150 and 300 K yields $C = 4.184(3)$ cm$^3$ K mol–Mn$^{-1}$ and $\Theta = 32.3(2)$ K, which are close to the reported values of $C = 4.1$ cm$^3$ K mol–Mn$^{-1}$ and $\Theta = 31.5$ K.[34] The positive Weiss temperature means that antiferromagnetic interactions are dominant and the Curie constant corresponds to an effective magnetic moment of 5.79 $\mu_B$ per Mn atom, which is reasonably close to a spin-only value of 5.92 $\mu_B$ for $S = 5/2$. Note that magnetic susceptibility deviates from the Curie−Weiss law below ~50 K and shows a broad peak at approximately 14 K. The ratio of the peak temperature to the Weiss temperature $T_p(\chi)/\Theta$ is about 0.44. It is further noted that the reduction in magnetic susceptibility below the peak temperature is unusually large and is almost comparable to that of a spin 1/2 Heisenberg antiferromagnetic chain.[41] Since such a large reduction in magnetic susceptibility is absent in other kagome- or triangular-lattice antiferromagnets especially with classical spins,[5,19-20] it is probably attributed to the development of strong antiferromagnetic correlations that are characteristic of the magnetic frustration of the KT lattice.

Magnetic susceptibilities measured at low temperatures and under various magnetic fields are shown in the inset to Fig. 3. A kink is observed at 2.0 K at a magnetic field of 0.1 T, which indicates an antiferromagnetic long-range order. As magnetic field increases, it moves to lower temperatures and disappears at 5 T, suggesting that the order is suppressed by weak magnetic fields. The frustration factor $f$, which is the ratio of the Weiss temperature to the transition temperature expressed as $f = |\Theta|/T_N$ and is used to measure the strength of magnetic frustration, is 16 for $NaBa_2Mn_3F_{11}$. It is larger than those of kagome antiferromagnets: 10 and 14 for Fe- and Cr-jarosites,[19,20] respectively, and 9 for vesignieite.[25] This suggests that the magnetic frustration is crucial in $NaBa_2Mn_3F_{11}$.

The heat capacity divided by temperature, $C/T$, of $NaBa_2Mn_3F_{11}$ increases with decreasing temperature below ~20 K and shows a jump at 2.0 K indicating a second-order phase transition, as shown in Fig. 4(a). Since the two transition temperatures from magnetic susceptibility and heat capacity coincide with each other, we conclude that an antiferromagnetic long-range order of bulk nature sets in at 2.0 K. Applying a magnetic field of 7 T seems to suppress the transition, suggesting the fragile nature of the order, as already observed in magnetic susceptibility.

In order to extract the magnetic contribution to heat capacity, the lattice contribution has been estimated by fitting the high-temperature data, as mentioned below; a nonmagnetic reference compound is not available for $NaBa_2Mn_3F_{11}$. Provided that the lattice contribution is the sum of Debye-type and Einstein-type heat capacities, $C_D$ and $C_E$, respectively, $C/T$ above 50 K, where the magnetic heat capacity may be negligible, was fitted to the equation $C/T = 3R\{aC_D/T + (17/3 − a)C_E/T\}$, where $R$ is the gas constant. The results of the fitting is reasonably good, as shown in Fig. 4(a), giving the parameters $a = 2.97(5)$, $T_D = 200(3)$ K, and $T_E = 485(5)$ K, where $T_D$ and $T_E$ are the Debye and Einstein temperatures, respectively. The thus obtained magnetic heat capacity $C_m$ and magnetic entropy $S_m$ are shown in Fig. 4(b). $C_m$ grows below ~20 K and shows a broad peak at 9 K as shown in the inset of Fig. 4(b) before the magnetic order at 2.0 K. The ratio of the peak temperature to the Weiss temperature, $T_p(C_m)/\Theta$, is approximately 0.28. The broad peak at 9 K in magnetic heat capacity must be attributed to an entropy release associated with the development of short-range magnetic correlations, which may correspond to the broad peak in magnetic susceptibility at approximately 14 K in Fig. 3.

Interestingly, the magnetic entropy consumed below $T_N$ corresponds to only 13% of the total entropy expected for spin 5/2 ($R\ln 6 = 14.9$ J K$^{-1}$ mol−Mn$^{-1}$), as shown in Fig. 4(b). Thus, most magnetic entropy is released by the short-range magnetic correlations above $T_N$, indicating the importance of magnetic frustration in the KT lattice. The magnetic entropy at 50 K reaches 13.3 J K$^{-1}$ mol−Mn$^{-1}$, which corresponds to 90% of the expected total entropy. Since this small deficiency in released entropy could be mainly due to the rough estimation of the lattice contribution at high temperatures, there may be no residual entropy at $T = 0$, suggesting that the spins are fully ordered.

We discuss the magnitude and sign of magnetic interactions in NaBa$_2$Mn$_3$F$_{11}$ from the viewpoint of the crystal structure. They cannot be simply decided in terms of interatomic distances as the magnetic interactions should occur by superexchange interactions via F ions. Generally, the superexchange interaction depends on the angle formed by magnetic ions and bridging ligand ions: ferromagnetic and antiferromagnetic interactions are expected for 90º and 180º angles, respectively, and their magnitude is small in between.[42] The nearest-neighbor magnetic interaction in NaBa$_2$Mn$_3$F$_{11}$ $J_1$ results via both the Mn−F1−Mn and Mn−F2−Mn paths with bond angles of 85.3(2)/100.7(2) and 117.23(9)º, respectively, as shown in Fig. 1(c); the two kinds of angle for the former comes from the splitting of the F1 sites due to structural disorder.[34] The former can be either ferromagnetic or weakly antiferromagnetic, while the latter must be antiferromagnetic. Since the actual interaction comes from the sum of these two contributions, it is difficult to predict the sign and magnitude of $J_1$. In contrast, the next-nearest-neighbor coupling $J_2$ results via the Mn−F1−Mn paths with a bond angle of 173.6(3)º and is definitely expected to be antiferromagnetic. Although the Mn−Mn distance is larger for this path (5.409 Å) than for the nearest-neighbor paths (3.562 Å), one would expect a significantly large $J_2$ because of the nearly linear bond. From these arguments, either one of the two cases is likely: antiferromagnetic $J_1$ and $J_2$ ($J_1 > 0$ and $J_2 > 0$) or ferromagnetic $J_1$ and antiferromagnetic $J_2$ ($J_1 < 0$ and $J_2 > 0$). In the latter case, the magnitude of antiferromagnetic $J_2$ should be more than twice as large as that of ferromagnetic $J_1$: the relation $J_2 > 2|J_1|$ is required to obtain a positive Weiss temperature, as observed here; $\Theta = (4J_1 + 2J_2)S(S + 1)/3k_B$ for $\chi = C/(T + \Theta)$ in the mean field approximation.

Magnetic orders for classical Heisenberg spins in the KT lattice have been theoretically examined. We have searched for wave vectors with the lowest energy by analysis based on the Fourier transform of the exchange interactions.[43] A triangular lattice model with $J_1$ and $J_2$ shown in Fig. 2(c) is considered. The phase diagram obtained in the $J_1$–$J_2$ space is shown in Fig. 5(a), where the positive sign of $J$ means an antiferromagnetic interaction. 120° orders with $q = 0$ and $\sqrt{3} \times \sqrt{3}$-type are stable in the cases of $J_1 > 0$ and $J_2 > 0$ and of $J_1 > 0$ and $J_2 < 0$, respectively, just as in the conventional $J_1$–$J_2$ kagome model.[16] On the other hand, a ferromagnetic order is stable for $J_1 < 0$ and $J_2 < 0$. The most interesting region is found for $J_1 < 0$ and $J_2 > 0$, where the two interactions compete with each other and generate frustration. A commensurate order and an incommensurate order are found for $|J_1| < J_2$ and $|J_1|/2 < J_2 < |J_1|$, respectively. The former is characterized by a wave vector at the M point of the first Brillouin zone, as shown in Fig. 5(b). Analyses based on classical Monte–Carlo simulation reveal that it is a non-coplanar order with twelve sublattices which is identical to the cuboc order found in a conventional Heisenberg model of the $J_1$–$J_2$ kagome lattice with $J_1 < 0$ and $J_2 > |J_1|/3$.[17,18] On the other hand, the latter incommensurate order seems a unique non-coplanar order that has been found in neither the kagome nor triangular lattice model. Interestingly, the characteristic wave vector of this order spans the distance between the zone center and the M point as $J_2/J_1$ varies. Details of these non-coplanar phases will be reported elsewhere.

In order to obtain insight into the magnetic order of $NaBa_2Mn_3F_{11}$, magnetic susceptibility and heat capacity are calculated by classical Monte–Carlo simulations for various combinations of $J_1$ and $J_2$ and compared with the experimental ones. In each simulation, a peak is observed as a function of temperature, which may correspond to the broad peak observed in the experiments. For example, the calculated magnetic susceptibilities for $J_2/J_1 = −4$ and $−5$ are compared with the experimental data in Fig. 3. The former reproduces the experimental peak temperature well, while the latter reproduces the peak height better but not the peak temperature. We assume that the reproduction of the peak temperature is more important and thus take the former, because the magnitude of the magnetic susceptibility may be sensitive to small perturbations to the KT lattice model, such as an inter-plane coupling or a DM interaction. In addition, our classical calculations may fail to treat quantum fluctuations appropriately.

The ratios of the peak temperature to the Weiss temperature are shown as a function of $J_2/J_1$ for magnetic susceptibility and heat capacity in Figs. 5(c) and 5(d), respectively. Both the experimentally obtained values of $T_p(\chi)/\Theta \sim 0.43$ and $T_p(C_m)/\Theta \sim 0.28$ are simultaneously reproduced for either $J_2/J_1 = -0.8$ or $-4$, which correspond to the $\sqrt{3} \times \sqrt{3}$-type 120° order and the cuboc order, respectively. Then, $J_1$ and $J_2$ are estimated by taking into account the Weiss temperature of 32.3 K: $(J_1/k_B, J_2/k_B) = (4.5 \text{ K}, -3.5 \text{ K})$ and $(-2.8 \text{ K}, 11.1 \text{ K})$, respectively. We think that the latter is more likely because an antiferromagnetic $J_2$ is expected from the crystal structure, as mentioned above. Therefore, a cuboc order could be realized if the simple Heisenberg model on the KT lattice is appropriate for the present compound. The cuboc order has never been observed experimentally, although it has recently been reported that kapellasite, a candidate compound for the spin 1/2 kagome antiferromagnet, shows a magnetic correlation toward the cuboc order.[44] Although the results of our analyses suggest a preference for a cuboc order, the possibility of other exotic orders cannot be excluded, when additional interactions such as DM interactions (expected from the crystal structure) are large enough to modify the $J_1$–$J_2$ phase diagram. Neutron diffraction and NMR experiments are in progress to determine the magnetic structure of $NaBa_2Mn_3F_{11}$. Single-crystal growth experiments are also in progress.

In summary, we have shown that $NaBa_2Mn_3F_{11}$ is a model compound of a kagome−triangular lattice antiferromagnet that can interconnect the kagome and triangular lattices. Magnetic susceptibility and heat capacity measurements reveal an extensive short-range magnetic order below 14 K and a long-range order at 2.0 K. Considerations of the crystal structure and comparisons of the experimental data with the data obtained by classical Monte−Carlo simulation suggest that a cuboc order or other exotic orders are realized at ferromagnetic $J_1$ and antiferromagnetic $J_2$. We think that there is a great chance to observe a unique magnetic order in classical spin systems on the KT lattice. It would also be interesting to examine how quantum fluctuations affect the ground state, resulting in an exotic order, in a quantum spin system on the KT lattice.

**Acknowledgments**

We thank K. Nawa, S. Hayashida, T. Masuda, G. J. Nilsen, M. Akaki, and M. Tokunaga for fruitful discussions. Part of this work was supported by the Strategic Programs for Innovative Research (SPIRE), MEXT and by the Computational Materials Science Initiative (CMSI), Japan. Some computations have been done using the facilities of the Supercomputer Center, the Institute for Solid State Physics, the University of Tokyo.

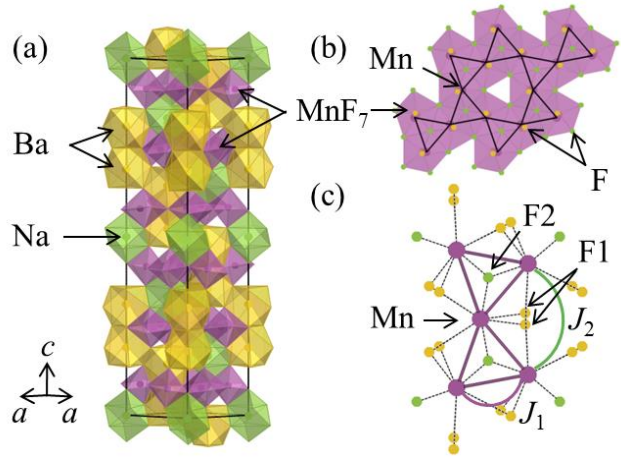

Fig. 1. (color online) (a) Crystal structure of $NaBa_2Mn_3F_{11}$ comprising layers composed of $MnF_7$ pentagonal bipyramids separated by $NaF_8$ and $BaF_9$ coordination polyhedra. (b) $Mn_3F_{11}$ layer viewed along the $c$-axis. The purple spheres represent Mn atoms that form a two-dimensional kagome–triangular lattice, as shown by the thick lines. The green spheres represent F atoms near the basal plane, and the yellow spheres are F atoms above Mn atoms. (c) Local atomic arrangements around a pair of Mn triangles. F1 atoms occupy one of the two adjacent sites randomly, and F2 atoms are located slightly above or below the Mn triangles. The nearest-neighbor magnetic interaction $J_1$ occurs by superexchange couplings via F1 [∠Mn–F–Mn = 85.3(2), 100.7(2)°] and F2 atoms [117.23(9)°], while the second-nearest-neighbor coupling $J_2$ occurs via F1 atoms [173.6(3)°]. The former can be either ferromagnetic or antiferromagnetic, while the latter must be antiferromagnetic.

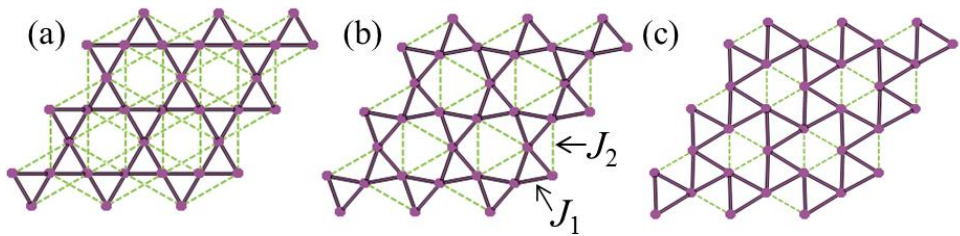

Fig. 2. (color online) (a) Kagome lattice with the nearest-neighbor interaction $J_1$ (solid line) and the next-neighbor interaction $J_2$ (broken line). (b) Kagome–triangular lattice with $J_1$ and half of the $J_2$ couplings compared with the $J_1$–$J_2$ kagome lattice. (c) Another representation of the kagome–triangular lattice, emphasizing its same topology as the triangular lattice; they are identical to each other for $J_1 = J_2$.

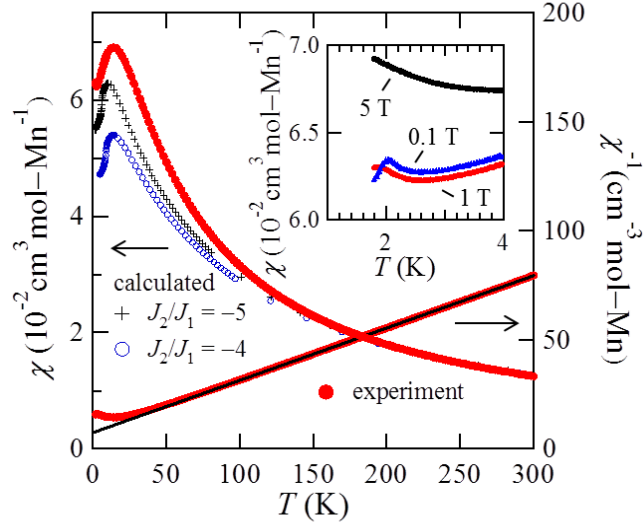

Fig. 3. (color online) Temperature dependences of magnetic susceptibility $\chi$ and its inverse measured between 1.8 and 300 K at a magnetic field of 1 T; both the zero-field-cooling and field-cooling processes are shown, which completely overlap to each other at any temperature. The magnetic susceptibilities calculated by classical Monte–Carlo simulation for $J_2/J_1 = -4$ and $-5$ with ferromagnetic $J_1$ and antiferromagnetic $J_2$ are also shown: $(J_1/k_B, J_2/k_B) = (-2.8$ K, 11.1 K) and $(-1.8$ K, 9.2 K), respectively. The solid line on the $\chi^{-1}$ data represents a Curie–Weiss fit, which gives an antiferromagnetic Weiss temperature of 32.3(2) K. The temperature dependences of the magnetic susceptibilities measured at low temperatures below 4 K at magnetic fields of 0.1, 1, and 5 T are shown in the inset. A kink indicative of a long-range order is observed at 2.0 K in the 0.1 T data.

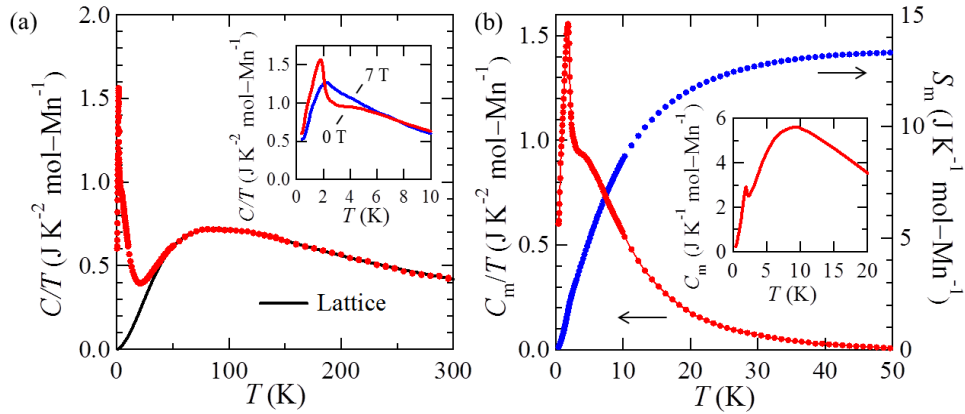

Fig. 4. (color online) (a) Temperature dependence of heat capacity $C/T$ measured between 0.4 and 300 K at zero magnetic field in the main panel and also at a magnetic field of 7 T in the inset. The solid line represents the lattice contribution estimated by fitting the data above 50 K, as described in the text. (b) Temperature dependences of magnetic heat capacity $C_m/T$ obtained by subtracting the lattice contributions of the total heat capacity and magnetic entropy $S_m$ obtained by integrating $C_m/T$. The temperature dependence of the magnetic heat capacity $C_m$ below 20 K is shown in the inset.

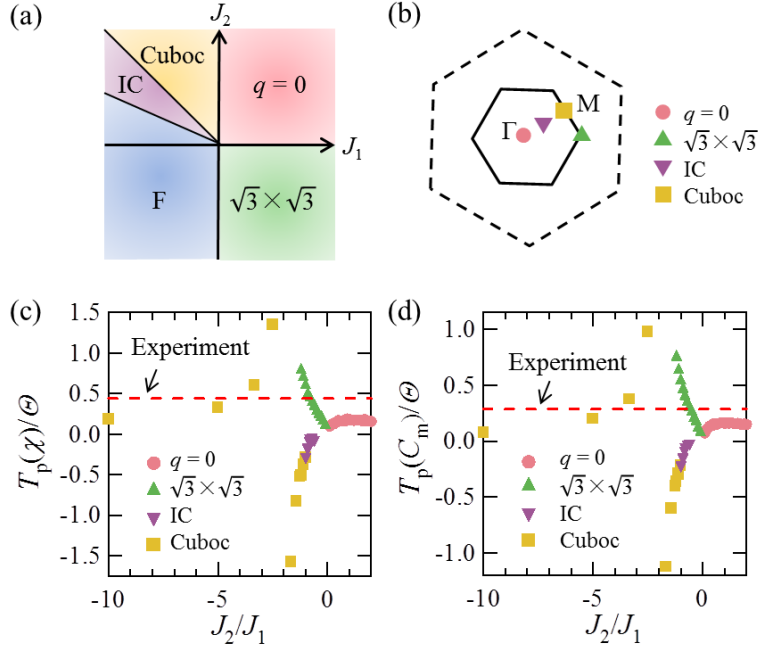

Fig. 5. (color online) (a) Magnetic phase diagram for classical Heisenberg spin in the kagome−triangular lattice. A positive sign of $J$ means an antiferromagnetic interaction. '$q = 0$' and '$\sqrt{3}\times\sqrt{3}$' refer to 120° spin orders with $q = 0$ and the $\sqrt{3}\times\sqrt{3}$-type on the kagome lattice, respectively, and 'F' refers to a ferromagnetic order. 'Cuboc' refers to a non-coplanar magnetic order with twelve sublattices, which is identical to that found in the $J_1$–$J_2$ kagome lattice model.[17,18] 'IC' is also a non-coplanar order with incommensurate wave vectors that may be unique for the KT lattice. (b) First Brillouin zones for the underlying triangular lattice (broken line) and the sublattice for the KT lattice (solid line) for the model shown in Fig. 2(c), where the characteristic wave vectors of the cuboc (square) and incommensurate (inverted triangle) are shown in addition to those of the 120° orders. The experimental ratios of the Weiss temperature to the peak temperature in magnetic susceptibility, $T_p(\chi)/\Theta$ (c), and in heat capacity, $T_p(C)/\Theta$ (d), are compared with the calculated values as a function of $J_2/J_1$ by classical Monte−Carlo simulation for an $L \times L$ triangular lattice under periodic boundary conditions ($L \leq 48$). The broken lines indicate the experimental values, which are successfully reproduced at $J_2/J_1 = -0.8$ for the $\sqrt{3}\times\sqrt{3}$ order and at $J_2/J_1 = -4$ for the cuboc order.